\documentclass[conference]{IEEEtran}
\usepackage{amsmath,amsfonts}
\usepackage{amssymb}
\usepackage{algorithmic}
\usepackage{algorithm}
\usepackage{array}
\usepackage{textcomp}
\usepackage{stfloats}
\usepackage{xurl}
\usepackage{verbatim}
\usepackage{graphicx}
\usepackage{cite}
\usepackage{subcaption}
\usepackage{multicol}
\usepackage{multirow}
\usepackage{xcolor}
\usepackage{pifont}
\usepackage{paralist}

\newcommand{\blu}[1]{\textcolor{black}{#1}}

\newcommand{\cmark}{\textcolor{blue}{\ding{51}}}%
\newcommand{\xmark}{\textcolor{red}{\ding{55}}}

\begin{document}

\title{Auditing the Grid-Based Placement of Private Label\\ Products on E-commerce Search Result Pages}

\author{\IEEEauthorblockN{Siddharth D Jaiswal}
\IEEEauthorblockA{\textit{IIT Kharagpur, India}}
\\
\IEEEauthorblockN{Yashwanth Babu Vunnam}
\IEEEauthorblockA{\textit{UM Amherst, USA}}
\and
\IEEEauthorblockN{Abhisek Dash}
\IEEEauthorblockA{\textit{MPI-SWS, Germany}}
\\
\IEEEauthorblockN{Saptarshi Ghosh}
\IEEEauthorblockA{\textit{IIT Kharagpur, India}}
\and 
\IEEEauthorblockN{Nitika Shroff}
\IEEEauthorblockA{\textit{JP Morgan, India}}
\\
\IEEEauthorblockN{Animesh Mukherjee}
\IEEEauthorblockA{\textit{IIT Kharagpur, India}}
}

\maketitle

\begin{abstract}
    E-commerce platforms support the needs and livelihoods of their two most important stakeholders -- customers and producers/sellers. Multiple algorithmic systems, like ``search'' systems mediate the interactions between these stakeholders by connecting customers to producers with relevant items. Search results include (i) private label (PL) products that are manufactured/sold by the platform itself, as well as (ii) third-party products on advertised / sponsored and \textit{organic} positions. In this paper, we systematically quantify the extent of PL product promotion on e-commerce search results for the two largest e-commerce platforms operating in India -- Amazon.in and Flipkart. By analyzing snapshots of search results across the two platforms, we discover high PL promotion on the initial result pages ($\approx 15\%$ PLs are advertised on the first SERP of Amazon). Both platforms use different strategies to promote their PL products, such as placing more PLs on the advertised positions -- while Amazon places them on the first, middle, and last rows of the search results, Flipkart places them on the first two positions and the (entire) last column of the search results. We discover that these product placement strategies of both platforms conform with existing user attention strategies proposed in the literature. Finally, to supplement the findings from the collected data, we conduct a survey among 68 participants on Amazon Mechanical Turk. The click pattern from our survey shows that users strongly prefer to click on products placed at positions that correspond to the PL products on the search results of Amazon, but not so strongly on Flipkart. The click-through rate follows previously proposed theoretically grounded user attention distribution patterns in a two-dimensional layout.
\end{abstract}
	
	%
 
    \section{Introduction}
\label{sec:intro}
E-commerce marketplaces, e.g., Amazon, Flipkart, Alibaba, etc., cater to the needs of two primary stakeholders -- customers and sellers of the marketplace. Customers depend on them for their purchase needs~\cite{un_ecommerce_growth} whereas sellers across the world rely on them for their livelihood~\cite{forbes2021smallsellers}. Due to their scale of operation, several algorithmic systems, e.g., search and recommendation systems, are deployed on these platforms for ease of accessibility of the desired products (and their sellers) to different customers. Traditionally, most of these algorithms are keyed to relevance and/or customer satisfaction~\cite{linden2003amazon, sorokina2016amazon, smith2017two}. 
However, realizing that these algorithms impact both customers and sellers, researchers have recently focused on fairness toward both stakeholders~\cite{singh2018fairness, mehrotra2017auditing, patro2020fairrec, dash2022fairir, mehrotra2018towards}. Most of the work in this space has focused on bias and fairness toward various demographic groups or toward an individual stakeholder. 
In contrast, our study focuses on an emerging strand of fairness research: \textit{potential preferential treatment of private label products (products manufactured by the marketplace platform) in the search results}. 
To this end, we conduct detailed experiments on two of the most prevalent e-commerce platforms in India -- Amazon.in~\cite{amazon_home} and Flipkart~\cite{flipkart_home}.
More precisely, the fairness concerns discussed in this work stem from (a)~interleaving sponsored results within organic search results and (b)~the over-promotion of private-label products using the search apparatus. Note that the above practices may significantly affect the mentioned stakeholders of the marketplaces. Next, we briefly discuss these practices and their fairness concerns.

\begin{figure*}[!t]
	\centering
	\centering
		\includegraphics[width=0.7\textwidth,height=0.7\textheight,keepaspectratio]{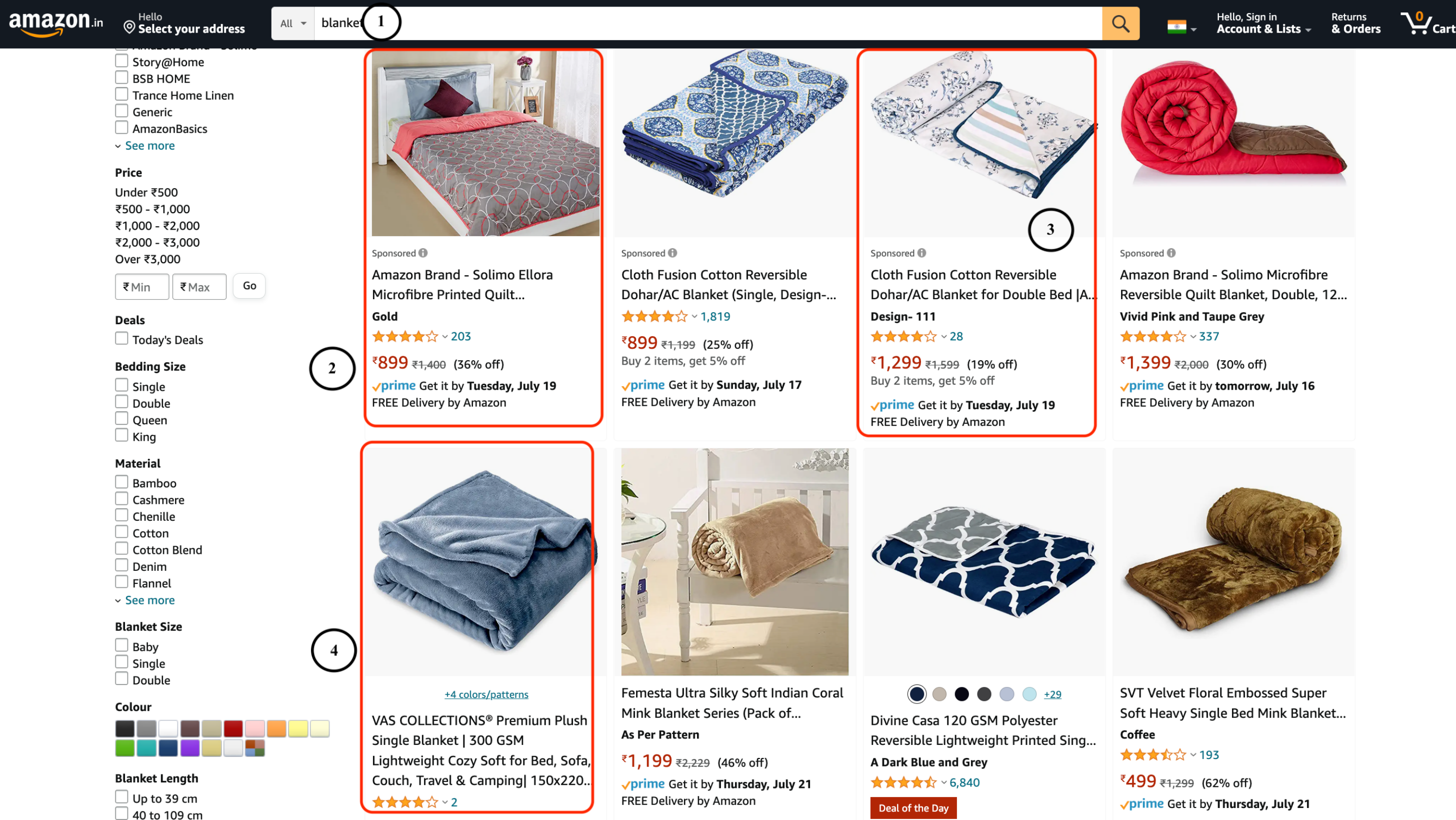}
	\vspace{-2 mm}
	\caption{\footnotesize{\bf First SERP for the Amazon e-commerce platform. The different markers are -- \textcircled{1} Search bar for typing in the search query, \textcircled{2} A typical private label (PL) product being shown in the sponsored position, \textcircled{3} A typical third-party (3P) product being shown in the sponsored position, \textcircled{4} A typical third-party (3P) product being shown in the organic position.}}
	\label{fig:search_ss}
\end{figure*}

\noindent {\bf Search results on e-commerce platforms}: \textit{Search engines} are one of the most popular ways to look for the desired products among customers~\cite{sorokina2016amazon}. Figure~\ref{fig:search_ss} shows a snapshot of such a search operation in Amazon. On these platforms, customers enter query words based on their requirements, and many products are displayed in decreasing order of relevance to the issued query on the search engine result pages (SERPs). Often, these results are also accompanied by some metadata regarding the product for better decision-making of the customers~\cite{dash2022alexa}. These search results are typically generated by optimizing for customer engagement signals such as click-through rate (CTR) or sales~\cite{sorokina2016amazon} and are frequently presented in a \textit{grid-based} layout. Traditionally, these search results are said to be \textit{organic} search results as they are derived from past customer behaviour.
Figure~\ref{fig:search_ss} shows the first search engine result page (SERP) for Amazon when the customer searches for ``blanket'' -- \textcircled{4} is the first organic result.\footnote{In some (rare) occasions, the results may be arranged in a single list instead of a grid. We do not consider this mode of arrangement in our study.}  

Due to the business model of the e-commerce platforms, increasingly advertised or \textit{sponsored results} are being interleaved amid the organic search results. These results are often decided based on \textit{second price auctions} for different queries. These are usually demarcated using a label: `Sponsored' on Amazon and `Ad' on Flipkart (see \textcircled{2} in Fig.~\ref{fig:search_ss}). Interestingly, sponsored search results are placed at fixed positions in the SERPs by default and often in very strategic positions. For example, on Amazon -- the first, middle and last rows of a SERP are advertisement positions, whereas on Flipkart -- the first two positions and the entire last column have advertised results. In both the platforms (Amazon and Flipkart), sponsored results precede the organic results on SERPs (Fig.~\ref{fig:search_ss}). This may alter the organic ranked list of products which is based on customers' prior preference. 

\noindent
\textbf{Customers interaction with e-commerce search interface}: Prior research in modelling users' interaction with web pages shows that results displayed in prominent positions on the SERPs receive more attention~\cite{baeza2018bias,singh2018fairness}. However, in many cases, customers do not distinguish between sponsored and organic results properly, leading to more clicks for some of the sponsored results~\cite{Graham2019Amazon}. Given sponsored results have monetary incentives associated with them, platforms attempt to put them in pronounced slots on the SERP conforming to standard user gaze patterns~\cite{eye_gaze_pattern1,eye_gaze_pattern2, xie2019grid}. 
Note that such practices impact the utility of both customers and sellers. Such smart positioning of sponsored products may steer customers away from organic products, which are more relevant based on prior customer interactions for the queries at hand. 
At the same time, the sellers of the organic products might be deprived of the opportunity to gain some revenue because the attention is being diverted to another seller who has outbid others to gain the top position of the search result (albeit sponsored). 

\noindent \textbf{Private label vs. third party products on e-commerce platforms}:
This brings us to the second facet of the argument, i.e., private label products. Generally, there are two types of brands on e-commerce platforms: (1)~private label brands and (2)~ third-party brands. The corresponding products from these brands are referred to as \textit{private label products} (PLs) and \textit{third party products} (3Ps), respectively. Private label products are produced and sold by the e-commerce platform itself. Amazon sells private label products under several private label brands such as -- AmazonBasics, Amazon Essentials, Pinzon, Presto etc.~\cite{amazon_pl}. Similarly, Flipkart 
sells products from private label brands such as -- MarQ, Perfect homes, Smart-Buy, Adrenex, etc.~\cite{flipkart_pl} on their platform. On the other hand, 3P products of different third-party brands, e.g. Duracell, Adidas, etc., are sold by several sellers on these platforms.

Notice that products from these private label brands are exclusively sold on the corresponding e-commerce platforms~\cite{Edelman2019Amazon}. Hence, selling such products often comes with an additional monetary incentive for these platforms~\cite{dash2021umpire, Kalra2021Amazon2}. Another important factor that complicates the matter even further is that both PL and 3P products can appear as both sponsored and organic search results. In such cases, marketplace organizations act both as an umpire and a player~\cite{dash2021umpire}. As an umpire, they control the search algorithm deployed on the platform, and as a player, their own private label products can be placed in organic search results. Similarly, as an umpire, they act as an ad exchange to collect bids from different sellers for ad spaces and as a player, they also place the advertisements of their private label products as advertisers. This dual role-play further raises fairness concerns on such business practices as it may not be a level playing field between the e-commerce marketplace and its competing sellers, who rely on them for their revenue. To this end, several media articles have recently surfaced which, either with anecdotal evidence or internal documents, show how Amazon may have potentially rigged search algorithms and/or results to nudge customers toward its PL products~\cite{Kalra2021Amazon2, Creswell2018Amazon}. To further underline the seriousness of the issue, such anti-competitive practices have landed e-commerce platforms, e.g., Amazon, into many antitrust lawsuits across the globe~\cite{House2020Judiciary, amz_anti_trust, EU2020Antitrust, India2019Antitrust, FTC2023AmazonSue}. 

\noindent\textbf{Research gaps in the existing literature}: Modelling of user interaction with ranked lists has been studied extensively over the years, but users observing grid-based search results (commonly shown on e-commerce SERPs) follow different attention distribution patterns~\cite{xie2019grid}. To the best of our knowledge, such interactions and their different inadvertent consequences are not yet well studied in the literature.
The evaluation metrics for grid-based results are not the same as standard list-based results. Measurement of relevance, fairness, etc., are modified to adapt to this change in presentation style. Evaluating and auditing grid-based results might require different treatment of the problem. 
Although there have been several works on fairness and bias related issues on e-commerce~\cite{dash2021umpire, dash2022alexa, Yin2021Search}, preferential treatment or issues of biases toward a set of products on sponsored search results has not been studied before.
To bridge this gap, in this work, we study the impact of sponsorship-induced private label promotion on grid-based search results by examining the product placement based on standard user gaze patterns~\cite{eye_gaze_pattern1,eye_gaze_pattern2}.

\noindent
\textbf{Legal positioning of private label promotion and sponsored results}: 
Note that while competition laws in many countries prohibit marketplace monopoly by abuse of power, most of these laws were written before the digital revolution~\cite{Sherman1890Antitrust, EC2016TFEU, GOI2016CompetitionAct}. Similarly, private label promotion is not a new phenomenon either. It also happens in brick-and-mortar shops in the physical world. However, none of the aforementioned laws explicitly address the concerns of such promotion. Having said that, the importance and impact of such practices increase in the digital world even further, especially due to the scale and magnitude at which they operate. To this end, the newly enforced Digital Markets Act (in the European Union) addresses some of these concerns. In particular, Article 6.5 of DMA~\cite{EC2022DMARegulation} explicitly prohibits such practices by mentioning, \textit{``The gatekeeper~\cite{EC2023Gatekeeper} shall not treat more favourably, in ranking and related indexing and crawling, services and products offered by the gatekeeper itself than similar services or products of a third party.''}\footnote{In accordance with the DMA, the European Commission can designate online platforms as `gatekeepers' if they act as a gateway between business users and end users in relation to core platform services. E-commerce platform Amazon is one of the six designated gatekeepers~\cite{EC2023Gatekeeper}.}
Similarly, Recital 70 in DMA~\cite{EC2022DMARegulation} asks gatekeepers not to present results to customers ``in a \textit{non-neutral} manner, or using the structure, function or manner of operation of a user interface or a part thereof to subvert or impair user autonomy, decision-making, or choice.''-- which essentially can be manipulated by smart positioning of private labels as sponsored results on SERPs.
On the other hand, since private label products are directly in competition with other third-party products, the extent of promotion of PLs in e-commerce search results has been investigated by authorities. In a recent investigation, Amazon claimed that only 2-3\% of the ad space is allotted to private label products~\cite{amz_anti_trust}. However, non-transparent positioning of PL products in the ad space can also potentially increase the price of the remaining ad estate, which may affect the third-party advertisers~\cite{amz_anti_trust}. Therefore, we posit that preferential treatment (if any) should not go unnoticed, even in the sponsored results.

\noindent\textbf{The current study}: 
In this study, we systematically quantify the extent of private label (PL) product promotion in e-commerce search results through simple descriptive measures. We differentiate our work from existing audits, done primarily on list-based search results, by focusing on 
\textit{grid-based search results}. We study the two most prevalent e-commerce platforms in India -- Amazon.in~\cite{amazon_home} and Flipkart~\cite{flipkart_home}. We investigate the extent of private label promotion from two different aspects: (1)~quantitative comparison of the volume of private label product promotion on the SERPs of the two e-commerce platforms; (2)~analysis of the different strategies of the positioning of products on the SERPs. To summarise, our key contributions and observations are as follows: 

\noindent
$\bullet$ We collect search results for 101 popular query phrases from Amazon and Flipkart over ten days, \blu{giving us more than \textit{2000 temporal queries}}. We calculate simple, succinct descriptive statistics that quantify the extent of private label promotion on grid-based search results of e-commerce platforms. 

\noindent
$\bullet$ We observe that both platforms promote private label (PL) products in sponsored / advertised search results, with Amazon's promotion being significantly higher. Across all the SERP, we find 11.7\% of all ad spaces to be occupied by Amazon PLs, and the percentage goes as high as 15.16\% for the first SERP. This is more than $5\times$ of what has been claimed by Amazon in anti-trust hearings~\cite{amz_anti_trust}.

\noindent
$\bullet$ These advertisements are positioned in very prominent locations on the SERP, e.g., at the top of the SERP (preceding organic search results) or just before the pagination buttons at the bottom. Such positioning has the potential to provide significant exposure to the sponsored results on a two-dimensional grid layout as hypothesized in the literature~\cite{xie2019grid}.
 
\noindent 
$\bullet$ To understand the nudging prowess and effectiveness of Amazon and Flipkart's different private label positioning strategies, we survey 68 participants on Amazon MTurk. In this survey, we simulate the SERP of e-commerce search and ask respondents to explore and order (add to cart) products of their choice.
	
\noindent
$\bullet$ The ordering patterns by survey participants show that the user-preferred click positions are similar to user click patterns hypothesized for grid-based search results in literature~\cite{xie2019grid}. Specifically, the survey participants' ordering for Amazon's products follows the middle bias (MB) pattern, where they prefer the middle columns of any given row. Similarly, for Flipkart's products, the ordering pattern follows the slower decay (SD) pattern, where their attention doesn't reduce monotonically. These click patterns strongly correspond to the positions of the private label products on the SERPs esp. for Amazon. We note that the platforms perhaps exploit this knowledge of users' behaviour and place their PL products in such positions to drive more attention toward them. Through a controlled experiment, we observe that Amazon's nudging prowess is more effective than placing PL products randomly. In fact, placing products following Amazon's PL promotion strategy \textit{triples} the number of clicks received by PLs, compared to a random arrangement of PLs in the SERP. 

We believe that the methodologies discussed in this paper can contribute to the deliberation surrounding the operationalization of newly adopted policies for fair competition in digital marketplaces, which can interest researchers, regulators and marketplace organizations alike. 

    \begin{table*}[!t]
	\noindent
 \tiny
	\centering
	\begin{tabular}{|p{2 cm}|p{8 cm}|p{1 cm}|p{1 cm}|p{1 cm}|p{1 cm}|}
		\hline 
		\textbf{Prior Works} & \textbf{Primary goal} & \textbf{Temporal} & \textbf{Preferential} & \textbf{Multiple} & \textbf{Customer} \\
            \textbf{} & \textbf{} & \textbf{Snapshots} & \textbf{Treatment} & \textbf{Platforms} & \textbf{Interaction} \\
		\hline
            Dash et al.\cite{dash2021umpire} & Preferential treatment toward Amazon private label products in related item recommendations. & \xmark  &  \cmark & \xmark & \xmark \\
        \hline
            Yin \& Jeffries\cite{Yin2021Search} & Preferential treatment in product search toward Amazon Private Label products. & \xmark  &  \cmark & \xmark & \xmark \\ 
        \hline
            Jaiswal \& Mukherjee \cite{jaiswal2022marching} & Preferential focus on perceived gender rather than clothing item in visual search for non-binary clothing items. & \xmark & \xmark & \cmark & \cmark \\
		\hline
            Dash et al.\cite{dash2022alexa} & Two-sided fairness and interpretability issues of conversational voice search in e-commerce. & \cmark  &  \xmark & \xmark & \cmark\\
            \hline
        \hline \hline
            Current work & Preferential treatment toward private label products in product search in Amazon and Flipkart in the perspective of customer interaction with most prevalent SERP interface. & \cmark  &  \cmark & \cmark & \cmark\\
            \hline
	\end{tabular}	
	\caption{\textbf{\textcolor{black}{Prior audit works on different algorithmic systems of e-commerce marketplaces, and their fundamental differences with the current work. Our study analyses the search engine result pages of 20 different temporal snapshots across two of the most popular e-commerce platforms in India. We also conduct multiple user surveys to corroborate the evidence observed in the empirical analyses of the data collected in our work. None of the prior works touched upon all the aspects in their audit studies.}}}
	\label{Tab: PriorWorks}
\end{table*}

\section{Related Work}
\label{sec:relwork}
In this section, we discuss two domains that contextualize our current work -- (a)~algorithmic auditing \& (b)~result placement strategies in search results.

\noindent \textbf{Algorithmic auditing}:
Modern AI systems are built on top of advanced and complicated algorithms. 
This has led to many potential biases~\cite{mehrabi2021survey} that can have far-reaching consequences. Academic researchers perform third-party audits~\cite{sandvig2014auditing,metaxa2021auditing} to evaluate such AI systems-- internet search~\cite{mehrotra2017auditing,robertson2018auditing,robertson2019auditing,hu2019auditing,robertson2018auditing2,robertson2020websearcher}, social media platforms~\cite{ali2019discrimination,sapiezynski2019algorithms,venkatadri2019auditing}, Youtube~\cite{ribeiro2020auditing,hussein2020measuring}, recommendation platforms ~\cite{dash2019network,wilson2021building,dash2022fairir}, e-commerce platforms~\cite{suvarna2019handling,dash2021umpire,jaiswal2022marching, juneja2021vaccine}, voice assistants~\cite{dash2022alexa}, and face recognition systems~\cite{buolamwini2018gender,jaiswal2022two}. 

\noindent
\textbf{Differences with close prior works on auditing information retrieval systems on e-commerce platforms:} \textcolor{black}{Multiple studies have audited~\cite{dash2019network,dash2021umpire, Yin2021Search} the different algorithmic systems on e-commerce platforms. However, most of these studies have been done in the context of Amazon. 
To succinctly differentiate our work from the prior works, we look at four important aspects (a)~ whether multiple temporal snapshots are studied, (b)~whether preferential treatment of any kind is studied, (c)~whether the observations can be extended to multiple platforms, and finally (d)~whether customer engagement based analyses are a part of it.
While Dash et al.~\cite{dash2021umpire} quantified the biases toward Amazon private labels in related item recommendations their study did not consider multiple temporal snapshots. Nor did they perform any user study to see how customers interact with the different recommendations on Amazon product pages. Yin \& Jeffries~\cite{Yin2021Search} performed an interesting study on Amazon search and how private labels appear toward the top of search. However, they also did not consider if the observations hold across a period of time or is it an artifact of the data collected within a specific time frame. Much like the previous work, it also did not include any user interaction-related surveys. Although Dash et al.\cite{dash2022alexa} and Jaiswal \& Mukherjee~\cite{jaiswal2022marching} in their studies took care of the longitudinal and user interaction aspects, the primary goal in these studies were not pertaining to the investigation of preferential treatment. Our audit study is primarily different from existing ones in the following ways (see Table~\ref{Tab: PriorWorks} for a summary) -- (1) we audit the SERPs on multiple e-commerce platforms for preferential treatment toward PL products, (2) our audit is performed on a very diverse set of query categories over a period of twenty temporal snapshots, and (3) our survey is more extensive and immersive, taken by 68 Amazon MTurk \textit{workers} directly on a mock shopping website.} 
Our proposed audit strategy, which identifies the preference of e-commerce platforms toward private label (PL) products in the grid-based search engine result pages (SERPs), is the first of its kind, to our knowledge.

\noindent \textbf{Placement strategies in search results}:
Search results are arranged on the display screen in multiple formats -- list (like in Google's textual SERPs) or grid (like in Google's image SERPs). Each of these has its own merits and the user attention patterns change based on this format. Research has shown that the user is focused more toward the top left half of the screen, and the attention decreases monotonically for list-based results~\cite{baeza2018bias}, whereas the user attention peaks in the middle of a row and changes non-monotonically between rows for grid-based results~\cite{xie2019grid,guo2020debiasing}. 
Till date, there has not been any effort to link the placement of products on e-commerce SERPs with user-attention patterns, as most existing studies have been performed on standard grid-based placement for image search. E-commerce search results have multiple metadata elements for each image in the search results, which may impact the attention pattern, and hence, existing observations and metrics may not translate directly. 

In this paper, we attempt to operationalize the preference of e-commerce platforms toward their private label (PL) products on the initial search engine result pages (SERPs), not just in terms of count but also in terms of the positions these products occupy on a grid-based layout in the SERPs. Our observations correlate with theoretical propositions on user preferences toward particular locations on the SERPs using grid-based layout. To the best of our knowledge, this work is the first such audit of its type on the Amazon platform and, any type of audit on the Flipkart platform. 

    \section{Dataset}
\label{sec:dataset}

\begin{table*}[!t]
    \scriptsize
	\begin{center}
		\begin{tabular}{| c | c | c | c | c | c | c | c | c | c |}
		    \hline
		    \multirow{3}{*}{\textbf{Query Category}} & \multirow{3}{*}{\textbf{Sample Queries}} & \multicolumn{3}{c}{\textbf{Amazon}} & \multicolumn{3}{|c|}{\textbf{Flipkart}}\\
		    \cline{3-8}
			&  & \multirow{2}{*}{\textbf{Queries}} & \multicolumn{2}{c|}{\textbf{ Products}} & \multirow{2}{*}{\textbf{Queries}} & \multicolumn{2}{c|}{\textbf{Products}} \\
			\cline{4-5}\cline{7-8}
			& & & \textbf{PL} & \textbf{3P} &  & \textbf{PL} & \textbf{3P}\\
			\hline
		    Electronics & trimmer, headphones, battery & 2 & 11 & 768 & 11 & 263 & 10695 \\ 
			\hline
		    Clothes \& Apparel & jeans, sweaters, backpack & 14 & 327 & 5042 & 16 & 602 & 15338\\
			\hline
			Home Appliances & fan, TV stand, kettle & 4 & 33 & 1524 & 3 & 43 & 2953 \\  
			\hline
			Sports & yoga mat, cycles, sports shoes & 5 & 112 & 1837 & 7 & 230 & 6721\\
			\hline
			Furniture & table, desk, shoe rack & 3 & 19 & 1144 & 9 & 421 & 8549\\
			\hline
			Cleaning Supplies & cleaning gloves, towel, detergent & 7 & 144 & 2579 & 2 & 10 & 1980\\
			\hline 
			Kitchen & coffee mug, water bottle, flask & 3 & 71 & 1088 & 5 & 78 & 4904\\
			\hline
			Home Decor & wall clock, bean bags, pillows & 3 & 110 & 1050 & 3 & 46 & 2944\\
			\hline
			Kids & baby footwear, baby innerwear, toy guns & $\varnothing$ & $\varnothing$ & $\varnothing$ & 4 & 282 & 3706\\
			\hline\hline
			\multicolumn{2}{|c|}{Total} & 41 & 827 & 15032 & 60 & 1975 & 57790\\
			\hline
		\end{tabular}
	\end{center}
	\caption{\footnotesize{\bf \blu{List of queries for which data is collected from each platform. Clothing \& Apparel is the category with the largest number of queries (and products) and Home Decor has the least. The total number of distinct queries is 78; 23 queries are common. The total number of products collected from Amazon is 15,859 and from Flipkart is 59,765.}}}
	\label{tab:queries}
\end{table*}

In this section, we describe the dataset collected from the Amazon and Flipkart queries as well as the process of collecting and cleaning it. We describe here the automated process of collecting the queries and their associated results. The data for this study was collected from both Amazon and Flipkart using the following steps -- (1) We first prepared a list of popular queries by searching for ``common'' or ``most searched'' queries on the Internet in the context of Amazon~\cite{amazon_query_src_1} and Flipkart~\cite{flipkart_query_src_1,flipkart_query_src_2,flipkart_query_src_3} platforms. These queries belong to various categories like Electronics, Fashion, Home Appliances, etc. and are diverse in terms of type, price and customer base. A total of \blu{101} queries were collected, out of which \blu{23} were common, giving us \blu{78} distinct queries on the two platforms across the various categories. Details regarding the categories, sample queries and number of queries per category are present in Table~\ref{tab:queries}. (2) Next, we used the Selenium web automation library to collect the results for each query. We performed the following steps to collect the data for each query in a sequential manner-- \textit{(a)} Open the website -- \url{https://www.amazon.in/} or \url{https://www.flipkart.com/} in a new \textcolor{black}{incognito} browser window; \textcolor{black}{no cookies are saved and the IPs stay the same}. \textit{(b)} Search for the query term without logging in, to avoid \textit{personalization} effects. \textit{(c)} Save the following information for every product in the result set - name of product, name of seller, unique product identifier, image, price and ratings. (3) The above process was performed twice a day with a gap of 8 hours between every two snapshots, for a period of 10 days, \blu{thus giving 2,020 \textit{temporal snapshots}}. This was done to ensure that the observations are not skewed due to the time and day of collection, and that the results could be generalized. Similarly, multiple snapshots and observations across them reduces the chance of stochasticity in the results. On Amazon, we have -- 41 queries (60 results per page); on Flipkart, we have -- 60 queries (40 results per page). 
The number of ads on the search engine result pages (SERP) always remains stable and does not vary based on any parameter, including browsing history.

    \section{Private label promotion on \\Amazon and Flipkart SERPs}
\label{sec:results}
In this section, we present our observations on the search results of the two platforms based on the dataset that we collected as discussed previously.
Our analyses aim to answer the following  questions: (1)~what fraction of private label products appear as sponsored results in the initial SERPs? 
(2)~how prominently are they located on these SERPs? 

\subsection{What fraction of PL products appear as sponsored results in the initial SERPs?}
\label{subsec:pl_in_ad}

\begin{table}[!t]
    \small
	\begin{center}
		\begin{tabular}{| c | c | c | c |} 
		\hline
		\textbf{Platform} & \textbf{SERPs} & \textbf{Total Results} & \textbf{\% PL in Ads} \\
		\hline
		\multirow{3}{*}{Amazon} & 1 & 2460 & \textbf{15.16}\\
        \cline{2-4}
         & 1 + 2 & 4920 & 13.98\\
        \cline{2-4}
         & 1 + 2 + 3 & 7380 & 13.10\\
        \hline \hline
        \multirow{3}{*}{Flipkart} & 1 & 2400 & 4.51 \\
        \cline{2-4}
         & 1 + 2 & 4800 & 4.47\\
        \cline{2-4}
         & 1 + 2 + 3 & 7200 & \textbf{4.56}\\
        \hline
		\end{tabular}
	\end{center}
	\caption{\footnotesize {\bf Percentage of Private Label (PL) products amongst all advertised products on the first three SERPS for both Amazon and Flipkart. \% PL in Ads indicates what \% of advertised products are PL. The values are decreasing from SERP 1 to 3. Maximum values are in bold.}}
	\label{tab:pl_in_ad}
\end{table}
We first analyze the ratio of Private Label (PL) products to the total advertised products in the \textit{first three SERPs}, for both Amazon and Flipkart. 
Note that both Amazon and Flipkart provide a larger number of SERPs (usually 7 for Amazon and 25 for Flipkart); however, majority of the customers never view all the search results, with more than 70\% not scrolling beyond Page 1~\cite{Rubin2019Cracking}. Hence, we focus on the first three SERPs. \textcolor{black}{Moreover, we observed similar trends for the later pages, hence we omit the results for brevity.} 

In Table~\ref{tab:pl_in_ad}, we present the \% of PL products among all advertised products on the first 3 SERPs (\% PL in Ads). This measure quantifies the \% of all ad-space available in the respective pages (cumulatively) that was allotted to private label products. We report the mean value for the 20 snapshots in our dataset. 

\noindent
\textbf{Observations}: We notice similar trends on both platforms -- the \% of PLs in advertisements is decreasing (non-increasing) from the first SERP to the first two SERPs to the first three SERPs (cumulatively). 
Interestingly, from Table~\ref{tab:pl_in_ad}, we observe that for Amazon, \blu{$\approx 15\%$} of the advertised positions are occupied by PLs on the first SERP, while for Flipkart this value is \blu{4.51\%}. This shows how different these platforms are even when they are both designed for the same task-- e-commerce. This justifies our decision of choosing two different platforms for our study.

Note that, in an anti-trust hearing~\cite{amz_anti_trust}, the committee asked Amazon -- \textit{``For each month since July 2018, please identify the percentage of all space eligible for advertising on Amazon that has been devoted to Amazon’s private label products.''} To this, Amazon responded that PL products take up 2--3\% of the total ad-space. 
In contrast, we observe that this percentage is 11.7\% considering all the search pages in our dataset. While we acknowledge that our dataset is limited to SERPs of only 41 queries, the differences in the percentages are noteworthy. Second, only the e-commerce platforms can answer what fraction of the entire ad space is provided to their private labels. However, this observation is important because 70\% of the shoppers never go beyond the first SERP~\cite{Rubin2019Cracking}. So, while the number may dilute when it is calculated across all available ad estate, its distribution in pages that are frequently visited by customers is of utmost importance.

We also calculate the `\% PL in Ads' per query category for the first SERP. For queries on Amazon in `Home Decor' category, PLs occupy $\approx 26\%$ of the ad space whereas for `Furniture' category, this is 2.9\%. Similarly, for Flipkart, the query category with the highest `\% PL in Ads' is `Kids' with 14.3\% and the lowest is `Cleaning Supplies' with 0\%.

Thus, we make the following observations from Table~\ref{tab:pl_in_ad}. 
\begin{compactitem}    
    \item Both Amazon and Flipkart promote PL products in sponsored / advertised search result on the initial SERPs. 
    Thus, even though both platforms are designed for the same task of e-commerce, the extent of promotion in Amazon is way higher than Flipkart.   
    \item These observations consistently appear across SERPs for queries of different categories.
\end{compactitem}

\subsection{How prominently are the PLs positioned in the initial SERPs?}
\label{subsec:position}

\begin{figure}[!t]
	\centering
	\begin{subfigure}{0.24\textwidth}
	\centering
		\includegraphics[height=4.5cm, keepaspectratio]{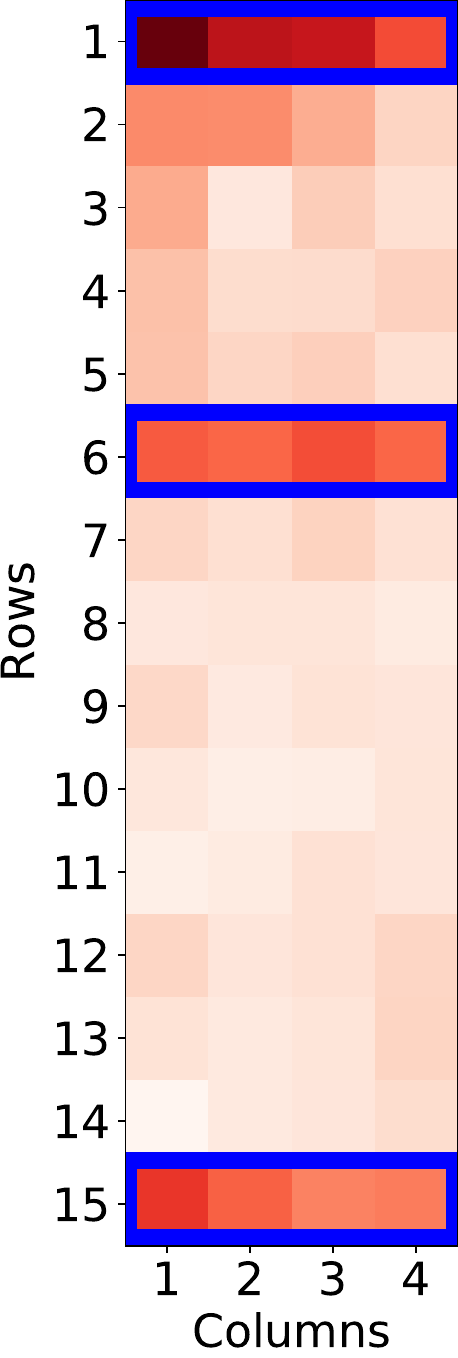}
		\caption{\footnotesize Grid results for \textsc{Amazon}}
		\label{fig:amz_pl_ad_matrix}
	\end{subfigure}%
	\centering
	\begin{subfigure}{0.24\textwidth}
	\centering
		\includegraphics[height=4.5cm, keepaspectratio]{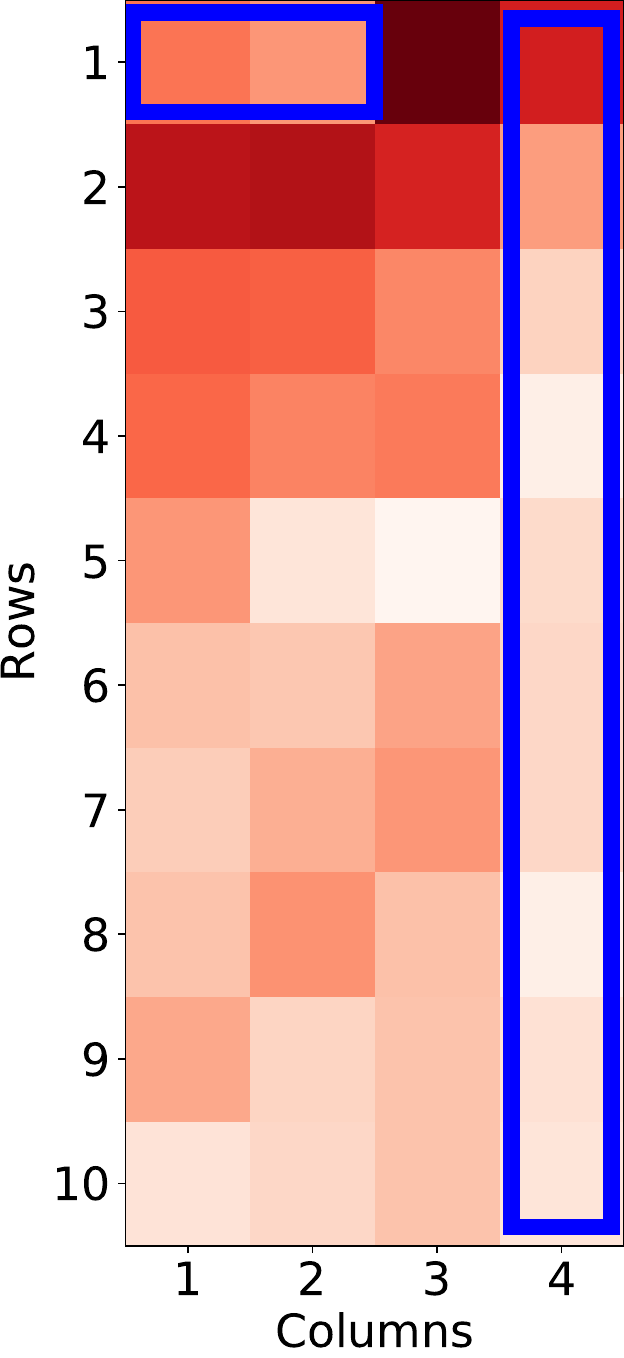}
		\caption{\footnotesize Grid results for \textsc{Flipkart}}
		\label{fig:fk_pl_ad_matrix}
	\end{subfigure}%
	\caption{\footnotesize{\bf Counts of the Private Label (PL) products and advertisement positions (highlighted by blue rectangles) on the first SERP of the Amazon and Flipkart platforms. PL products are present on the entire page, irrespective of the platform. On Amazon, the maximum occurrences of PL products are on the advertisement positions. On Flipkart, a higher count of PL products is shown in the top half (first 4 rows) of the SERP.}}
	\label{fig:pl_pos_ad_pos}
	\vspace*{-5 mm}
\end{figure}

Now that we have seen both Amazon and Flipkart placing PL products on the initial SERPs, we next look at the positioning of these products. 
That the position of a product/result in a SERP determines how much attention / exposure it receives from users, is a well-studied phenomenon~\cite{craswell2008positionbias, dash2021umpire, dash2022alexa}. 
We now analyse the positions in the SERPs where the PLs are displayed.

The heatmaps in Figure~\ref{fig:pl_pos_ad_pos} shows the positioning of the search results for both Amazon and Flipkart.
This figure shows the frequency of PL product appearance aggregated for all queries across the 20 snapshots at the different positions on the SERP grid. The intensity of color on the heatmaps indicates the count of PL products at a given position and the blue rectangular boxes indicate the positions reserved for advertisements on the SERP. 

\noindent
\textbf{Observations on Amazon}: For the Amazon platform, we observe that the PL products are distributed across the entire SERP grid; however, the intensity reduces as we go down the grid. Interestingly, in Fig~\ref{fig:amz_pl_ad_matrix}, we see that significantly higher counts are observed in rows 1, 6 and 15, \blu{which are already reserved for advertisements}. 
While Amazon may claim that it forgoes the ad revenue that it may have collected from the said positions (if they had shown a sponsored third-party product)~\cite{amz_anti_trust}, many media articles and experts provide a counterargument of how the exposure gained for PL products may out-weigh the short-term loss in ad revenue~\cite{propublica_pl_pref, dash2021umpire}.
Since the first 6 rows already have a higher count of PL products than the lower 9 rows, the aggregate attention received by the PLs would also be high. 
Thus, Amazon uses its sponsored result positions significantly for promoting its private label products as compared to the organic positions on the SERP. This, in turn, may directly contribute to increased exposure and revenue for the PL products on the platform. 

\noindent
\textbf{Observations on Flipkart}: Taking a look at the heatmaps for the Flipkart platform in Fig.~\ref{fig:fk_pl_ad_matrix}, we see that in the grid representation, the PL products are clustered in the top half of the page (rows 1-4) similar to Amazon.
However, unlike Amazon, many of these PLs are shown in organic search result positions. 
The advertised positions (marked with blue rectangles) are very different - the first two positions in row 1 (i.e., positions 1 and 2 of the SERP), and column 4. While private label positioning on Flipkart is still top-heavy, they appear more often on organic positions rather than on sponsored result positions.

\noindent
\textbf{Key differences in positioning of PLs by Amazon and Flipkart}:  Amazon is very systematic in PL advertising and positioning of PLs in the advertisement spaces. For example, Amazon's ad positioning covers the first row entirely which gathers the most exposure; followed by middle row and the final row (above the pagination buttons). Such positions indeed provide extra exposure to products appearing there and the same is also in line with some of the popular eye gazing attention patterns in the literature~\cite{eye_gaze_pattern1,eye_gaze_pattern2}.
In contrast, Flipkart's ad and PL placement strategies are very different. Flipkart reserves the top-left corner of the SERP for ads, much like Amazon does. However, the similarity ends there, \textcolor{black}{another evidence that necessitates the study of multiple platforms}. 
In the next section, we analyze the effectiveness of the two platforms' PL placement strategies to understand the benefits that they could possibly garner in terms of clicks and orders.
    \section{Survey to understand positional preferences}
\label{sec:survey}
The previous section shows how both platforms -- Amazon and Flipkart - place advertised search results on the search engine result pages (SERPs) in a grid layout and how they use this advertising infrastructure to promote their private label products. 
We now conduct a survey to investigate the two platforms for the following questions -- 
(Q1)~\textit{Do the two platforms' strategies of private label product placement in advertised search results have any nudging prowess for a customer?}, and (Q2)~\textit{Do the clicking patterns of users share any correlation with expected user attention patterns from literature, such as~\cite{xie2019grid}?} 
To answer these questions, we design a mock e-commerce website that simulates a real-world shopping experience, and get Amazon Mechanical Turk~\cite{amz_mturk} workers to participate in the study.

\subsection{Design of the survey}
\label{subsec:surveydesc}
\begin{table}[!t]
    \tiny
    \centering
    \begin{tabular}{|c|c||c|c|}
    \hline
        \multicolumn{2}{|c||}{\textbf{Amazon}} & \multicolumn{2}{c|}{\textbf{Flipkart}} \\ \hline
        Bean bag & Shower Curtain & Bean Bags & Table \\ \hline
        Blanket & Sweaters & Beds & Sweaters \\ \hline
        Desk & Table Lamp & Bedsheet & Toilet brush \\ \hline
        Dumbbells and Weights & Tissue paper & Cleaning Gloves & Power Bank \\ \hline
        Jeans & Trimmer & Coffee Mugs & Sports shoes \\ \hline
        Pans & Wall Clock & Cycles & Wall Clock \\ \hline
        Paper Towel & Water Bottle & Fan & Mobile covers \\ \hline
        Shoe Rack & Yoga Mat & Kettle & Lunch box \\ \hline
    \end{tabular}
    \caption{\footnotesize{\bf \textcolor{black}{Queries used in the survey for the Amazon and Flipkart platforms. For each query, we show the SERP as it appears on the original platform in the $W^O$ websites and a randomized version of the SERP in the $W^R$ websites.}}}
    \label{tab:survey_queries}
\end{table}

We create four mock shopping platforms using a free Wordpress~\cite{wordpress} website, hosted on individual AWS EC2 instances, with the open-source WooCommerce~\cite{woocommerce} e-commerce plugin, with a fixed set of queries. Two of the mock platforms show the first SERP for the queries exactly as they appear on the Amazon and Flipkart websites. The remaining two mock platforms show the first SERP's results in a randomized order. 
\textcolor{black}{Randomizing only within the advertised positions is not very meaningful as the number of such positions is limited and fixed. Hence we randomize the entire SERP's results, which provides a more meaningful baseline.}
We refer to the first set of platforms as $W_A^O$ and $W_F^O$ for Amazon and Flipkart respectively. Similarly, the platforms with the randomized arrangement of results are-- $W_A^R$ and $W_F^R$. 

To choose the queries for the mock websites, we first prune our dataset to a single snapshot.
From this, we randomly chose \textit{sixteen} representative queries (due to budget restrictions) as shown in Table~\ref{tab:survey_queries}. The same set of queries are present in both the $W^O$ and $W^R$ websites. Only results from the first search engine result page (SERP) was displayed to reduce the cognitive workload of the survey participant. $W_A^O$ and $W_A^R$ have 60 results (grid of 15 $\times$ 4, similar to Amazon's SERP) in the SERP for each query. Similarly, $W_F^O$ and $W_F^R$ have 40 results (grid of 10 $\times$ 4, similar to Flipkart's SERP) per SERP. Both websites had the same base template and differed only in terms of the queries and associated results. On the SERPs, we show some of the important intrinsic properties of every item -- \blu{product photos, names, brands, sizes, etc. However, we categorically avoid showing any ratings. We also avoid showing the ``Sponsored'' and ``Ad'' labels on the products to understand the participants' preference for the positions on the SERP irrespective of the label that the product has. These steps are meant to ensure that the survey participants’ choices are not biased toward certain products due to their historical preferences.}

\noindent \textbf{Steps of the survey:} The survey is performed as follows-- (1) The participant opens the landing page of the mock e-commerce website where they are informed of the goals of the survey and they consent to participate in the survey. (2) The second page shows the list of queries and the instructions that the participant must follow to complete the survey. (3) The instructions state that the participant must \textit{choose at least eight queries} of their choice and visit the results page for these. In the results page, they may choose to view a product by `\textit{clicking}' on its link or, `\textit{add}' it to cart. 
Both these actions are tracked in an aggregated manner to understand which products (thereby positions on the SERP) are interesting for the shopper. 
(4)~Once the participant has added the products (at least one per query) of their choice to the cart, they can proceed to the checkout page where they enter their PII details
(these are never used for anything other than the aggregated demographic information) along with a short explanation for their choices. Following this, they may submit and end the survey. 

\blu{We recognize a limitation of our survey method that simply adding to cart does not imply real shopping behaviour. 
However, asking respondents to actually purchase products is not feasible for third-party academic researchers like ourselves, due to budgetary concerns. We believe our survey format is the closest one can get to real world A/B testing.}

\blu{Note that, for such surveys, our institute does not require an IRB approval unless there is a medical intervention into the respondent's body. We refrain from collecting and disclosing any personal sensitive information from the respondents. In addition, the goals of the survey are mentioned on both the MTurk survey page as well as the landing page of our websites and explicit consent is sought from the participants before the survey begins.}

Overall, 29 Amazon Mechanical Turk participants took the survey for the website with Amazon's queries and 39 participants took the survey for the website with Flipkart's queries. 
A majority of the respondents belong to the $18-25$ age group and are undergraduate students. 
$\approx 43\%$ of them identify themselves as male and $\approx 45\%$ identify themselves as females; $\approx 12\%$  candidates opt not to disclose their gender.
The maximum time allowed for the survey was 30 minutes on Amazon Mechanical Turk (MTurk), and each worker was remunerated with \$0.50 if they successfully completed the survey by following the instructions described above. The average time taken per worker was $\approx 14$ minutes.

\subsection{Observations from order patterns in the survey}
\label{subsec:survey_obsv}

\begin{figure}[!t]
	\centering
	\begin{subfigure}{0.24\textwidth}
	\centering
		\includegraphics[height=4.5cm, keepaspectratio]{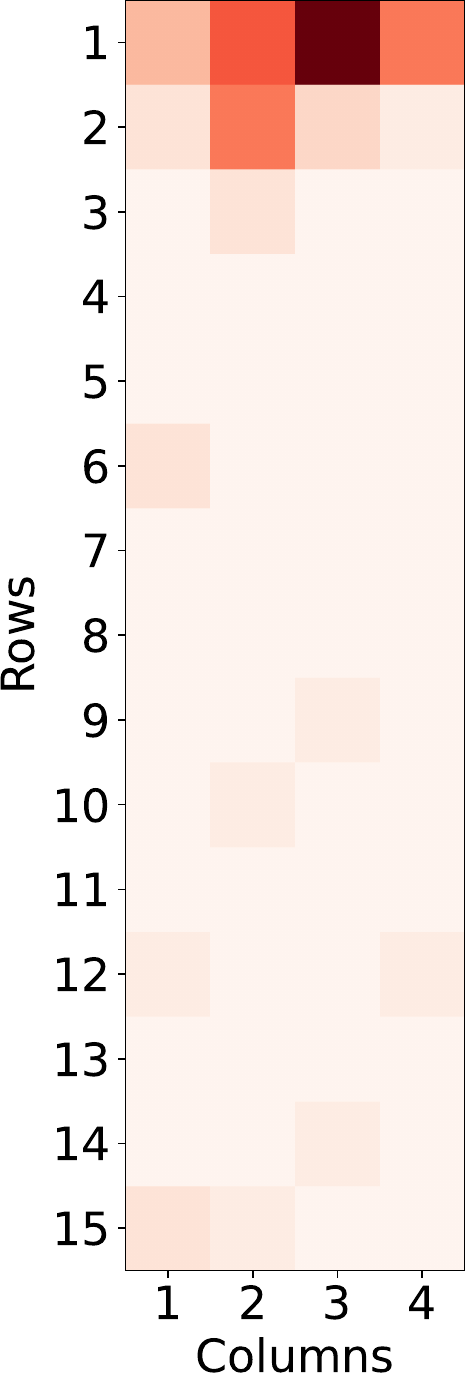}
		\caption{\footnotesize Orders for \textsc{Amazon}}
		\label{fig:amz_order_matrix}
	\end{subfigure}%
	~\begin{subfigure}{0.24\textwidth}
    \centering		
		\includegraphics[height=4.5cm, keepaspectratio]{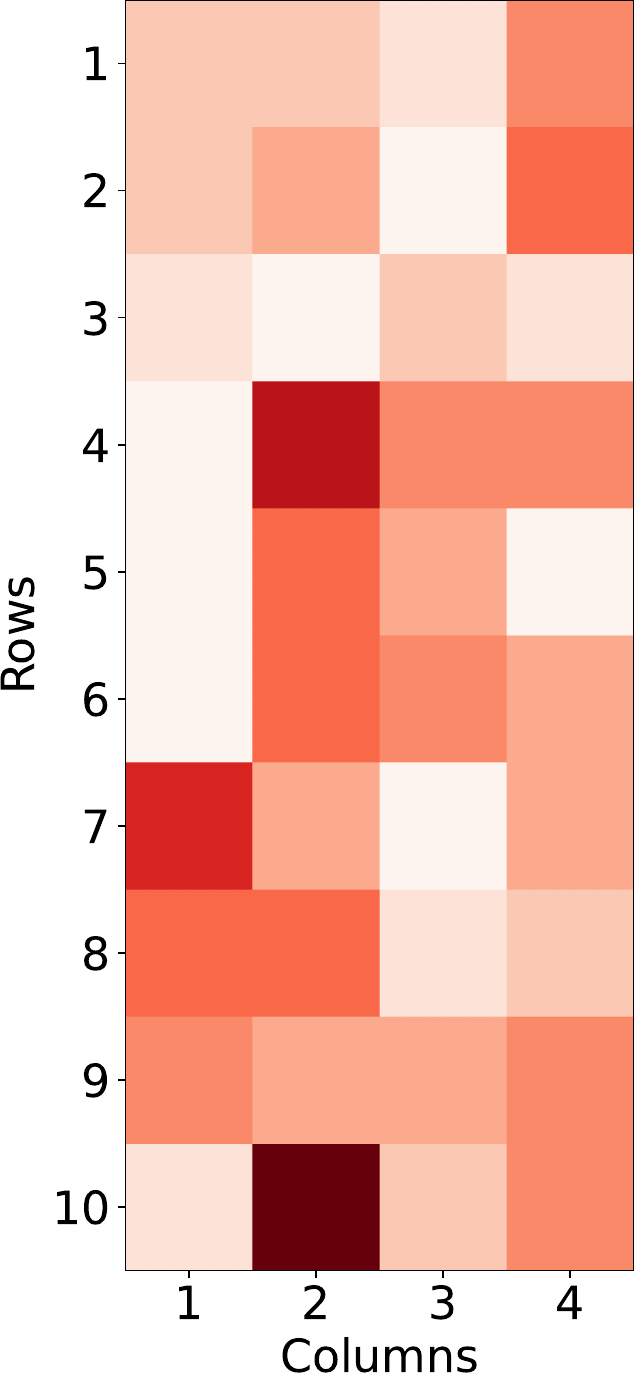}
		\caption{\footnotesize Orders for \textsc{Flipkart}}
		\label{fig:fk_order_matrix}
	\end{subfigure}
	\vspace{-2 mm}
	\caption{\footnotesize{\bf \textcolor{black}{Orders for all queries on the websites with original SERPs for Amazon ($W_A^O$) and Flipkart ($W_F^O$). For Amazon, the orders are clustered toward the top whereas they are distributed throughout the SERP for Flipkart, but with a higher ratio on the last column. This indicates that placing PLs at such positions can lead to more attention and orders for the same. These observation also falls in line with our observation of PL product placement in Figure~\ref{fig:pl_pos_ad_pos}}.}}
	\label{fig:order_matrix}
\end{figure}

\textcolor{black}{We record the orders of the respondents in the survey. In Figure~\ref{fig:order_matrix} we compare the orders received for products at different positions for all queries on each platform's survey. The intensity of the color of each cell corresponds to the number of orders the product at the position had gotten during our user survey.}

\noindent \textit{Insights from order patterns of respondents}: 
\textcolor{black}{From Fig.~\ref{fig:amz_order_matrix} we can see that a majority of the participants' clicks (and therefore their attention) is focused on the first three rows of the search results. This observation is in line with multiple prior observation on biases that appear on web-page browsing. The observations from Fig.~\ref{fig:fk_order_matrix} shows that the users have a preference toward the last column and bottom half of the SERP. }

\noindent
\textcolor{black}{\textit{Positions getting more orders $\equiv$ positions where PLs get placed more?}: Interestingly, the order patterns on the individual heatmaps correlate with the observed advertised PL positioning from Fig.~\ref{fig:amz_pl_ad_matrix}. 
For instance, the maximum number of orders in Fig.~\ref{fig:amz_order_matrix} are on the first row and second row.
This is similar to the placement of advertisement slots and dense private label positioning on Amazon SERPs (see Fig.~\ref{fig:amz_pl_ad_matrix}). 
On the other hand, the order pattern from Fig.~\ref{fig:fk_order_matrix} are focused on the bottom half of the SERP, mostly the middle columns, deviating from the placement position of the private label products as seen in Fig.~\ref{fig:fk_pl_ad_matrix}.} 

Thus we are able to successfully answer our first question -- \textit{The e-commerce platform's strategy of private label product placement in advertised search results can have nudging prowess for a customer}, as observed from the orders pattern on both the websites. However, Amazon's ad positioning, especially on the first row of the SERP has the most significant effect on customer behavior as compared to Flipkart's positioning on the right most column. 

\noindent
\textcolor{black}{\textit{Correlation with user attention patterns from literature}: To answer the second question, we calculate the correlation between the order pattern in our user surveys and the theoretical user attention as per~\cite{xie2019grid}. Even though e-commerce search results are not equivalent to image search results, they share some similarities -- presence of large images of the products and grid-like search result placement. We know that there exist multiple attention patterns for grid based image search results that users may follow--}
\begin{compactitem}
    \item \textit{Middle Bias} (MB): Users may focus more on the items in the middle columns of a row than on the outer columns. The stopping probability, i.e., the probability that a user views a particular item in a given row and column is calculated as $ \prod_{j=0}^{i-1} f(c(i)) C_j(1-C_i)$, where $c(i)$ is the column of item at rank $i$, $f()$ is a function of the column, and $C_i$ is the continuation probability at position $i$ to examine $(i+1)^\textrm{th}$ result. From~\cite{xie2019grid}, $f(c(i)) = e^{g(c(i))}$, where $g(c(i)) = \frac{1}{\sqrt{2\pi \sigma^2}}e^{-\frac{(c(i)-MP-\mu)^2}{2\sigma^2}}$. Here $MP$ is the column index of the middle position of a given row $r(i)$.
    \item \textit{Slower Decay} (SD): User attention may not reduce monotonically and/or dramatically with the rank of items, especially across rows. The stopping probability is calculated as $ \prod_{j=0}^{i-1} \beta^{r(i)}C_j(1-C_i)$, where $\beta$ is a constant parameter and $r(i)$ is the row of item at rank $i$. 
\end{compactitem}
The continuation probability is $C_i = p$ for both the above attention patterns. To calculate the correlations, we do the following -- for each cell, 
\blu{we calculate the rank from the number of orders.}
Similarly, we use the formulae described above to calculate the rank of each cell based on the parameters listed below. Then we calculate the Pearson correlation between the two ranked lists to judge whether the survey participant clicking pattern is positively correlated to what has been hypothesized in the literature.

\textcolor{black}{For middle bias, we set the middle position (MP) = 2, $\sigma = 0.5$, $p = 0.8$. The list-based metric we use for our calculations is RBP~\cite{moffat2008rank}. 
For the survey on Amazon's queries, we observe a Pearson correlation coefficient of $0.403$ ($p$-value = 0.0013) and for the survey on Flipkart's queries, the correlation coefficient is $-0.260$ ($p$-value = 0.104). Thus, we see a positive correlation between the theoretically expected user attention and the one observed in our survey for Amazon and a negative one for Flipkart.}

\textcolor{black}{For slower decay, the parameters are  $\beta = 1.5$ and $p = 0.8$. Interestingly, here the Pearson correlation coefficients are $-0.087$ ($p$-value = 0.50) for Amazon, indicating a negative correlation, and $0.114$ ($p$-value = 0.48) for Flipkart, indicating a weak positive correlation.}

\textcolor{black}{The above results show that on Amazon, while the users in our survey are more inclined to click on the middle columns within a row (middle bias), their attention does not drop non-monotonically with a decrease in rank (no slower decay). The opposite is true for Flipkart. The non-correlation on Amazon with slower decay may be attributed to the following-- the model proposed by~\cite{xie2019grid} is defined for image search and is not directly applicable to e-commerce search due to the presence of other features along with the image like price, ratings, etc. Thus, we confirm that the observed clicking patterns for Amazon do indeed share a highly positive correlation with one of the expected user attention patterns from literature, i.e., middle bias (MB). Similarly, we see a positive correlation on the Flipkart website with the slower decay (SD) attention pattern.}

\begin{figure}[!t]
	\centering
		\includegraphics[height=3cm, keepaspectratio]{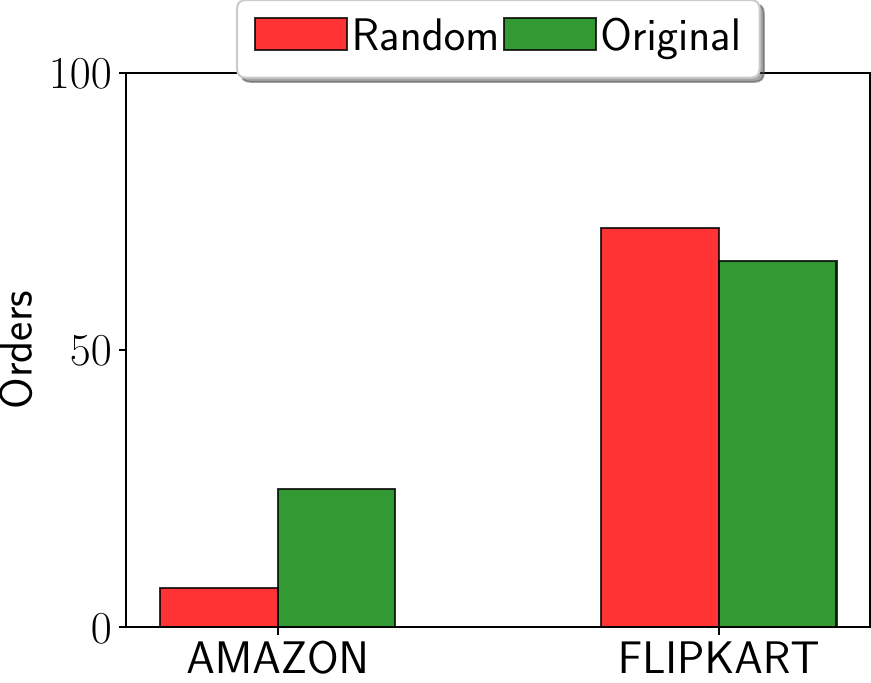}
	\caption{\footnotesize{\bf \textcolor{black}{Comparison of orders between queries where the PL products are placed randomly and where the PL products are placed as per the original distribution observed in the collected data. The uptick in number of clicks and orders for Amazon PLs further underpins the effectiveness of their PL promotion strategy. For Flipkart, the drop is marginal.}}}
	\label{fig:rand_orig}
	\vspace*{-3mm}
\end{figure}

\subsection{Comparing original against random PL placement}
\label{sec: effectiveStrategy}
Previously, we mentioned that for each platform, in one survey website, we show the results as they appear in the original SERP whereas in the other website, we randomize the order of the results. We now compare the counts of orders received by the PL products between the two sets in Figure~\ref{fig:rand_orig} to understand how effective are the two PL promotion strategies.
Note that, if we observe marginal or no differences in the volume of orders received by PL products in the random and original positioning setups, that brushes aside any concern of biases due to the corresponding promotion strategy.

\noindent
\textit{Insights from orders received by PLs}:
\textcolor{black}{We look at the comparison between the two sets of queries for orders in Figure~\ref{fig:rand_orig}. Here, we notice for Amazon that the survey participants click on PLs more often (upwards of $3\times$) if they are placed as per the original distribution (25 orders) obtained from the dataset rather than randomly (7 orders). For Flipkart, this trend doesn't hold true; infact participants prefer ordering products slightly more often (72 order) if they arranged as per a random distribution than the original distribution (66 orders).}

\textcolor{black}{Thus, from the results in Figs.~\ref{fig:order_matrix} and~\ref{fig:rand_orig} we see that Amazon reaps significant benefits from the positionining of the PL products as opposed to the Flipkart, where the positioning does not provide any discernible benefit. Not only are Amazon's PL products clicked more often, but the survey participants order them more often as well. This further demonstrates that the systematic and sophisticated private label product promotion on the Amazon e-commerce platform can have a direct impact on revenue while the same may not be so pronounced for Flipkart. }

From the clicking and ordering patterns of respondents in the conducted survey, we make the following observations.
(1)~The ordering (therefore purchase) pattern of survey participants falls in line with the PL product promotion (through advertisements) of the Amazon e-commerce platforms that we studied.
(2)~The clicking pattern of survey participants is strongly positively correlated to the user attention patterns described in the literature and both the platforms exploit this to position their PL products so that they garner the maximum attention. Amazon follows the Middle bias pattern while Flipkart follows the Slower decay pattern.
(3)~Through the controlled experiment, we observe that the nudging prowess of Amazon's PL product promotion is more effective than that of Flipkart.

\vspace{1mm}
\noindent
\textbf{Significance of the observations:} 
The fact that products appearing towards the top of the search results derive more exposure (e.g., clicks and orders) has been well established in the literature of search ranking~\cite{baeza2018bias,singh2018fairness}.
Currently e-commerce platforms are using these positions extensively for product advertisements. 
As shown earlier, the same positions are also dominated by their private label products (see Figure~\ref{fig:pl_pos_ad_pos}). 
So, had these positions \textit{not} been reserved by the private label product ads, the same exposure could have been distributed over other third party products (and their corresponding producers and sellers). 
This, in fact, is also corroborated by our controlled experiment (explained above) where the PLs obtain much higher exposure (orders) when they are strategically placed (following Amazon's strategies) than when they are placed randomly. 
Hence, the insights from the click patterns coupled with the empirical observations in the previous section further show that placing PL products strategically and giving them preferential exposure directly translates to an increase in attention and revenue for the said platform.

    \section{Discussion \& future work}
\label{sec:discussion}

\begin{table*}[t]
	\noindent
 \scriptsize
	\centering
	\begin{tabular}{|p{.85\columnwidth}|p{.85\columnwidth}|}
		\hline 
		\textbf{Legal questions and / or recommendations} & \textbf{Potential answers from the current study} \\
            \hline
            Gatekeepers should not engage in behaviour... Such behaviour includes the design used by the gatekeeper, the presentation of end-user choices in a non-neutral manner, or using the structure, function or manner of operation of a user interface or a part thereof to subvert or impair user autonomy, decision-making, or choice. &  E-commerce platforms put sponsored results (which also contain their private label products) at strategic positions on the SERP interface which appeal to customers the most as per prior research and the conducted survey in this study. $62.5\%$ ($W_A^O$) of all adding to cart or clicks happen from the first row of the Amazon SERP during our survey.\\ 
            \hline
            The gatekeeper shall not treat more favourably, in ranking and related indexing and crawling, services and products offered by the gatekeeper itself than similar services or products of a third party. 
            & The strategic positioning of the private label products make them $3\times$ ($W_A^O$ vs $W_A^R$) more likely to be added to cart or clicked as opposed to any random positioning of the same set of products (as per our survey). \\
            \hline
            Please identify the percentage of all space eligible for
advertising on Amazon that has been devoted to Amazon’s private label products. & We find 11.7\% of all ad-spaces analysed in our Amazon dataset to be occupied by
Amazon PLs and the percentage goes as high as 15.16\% for the ad spaces in first SERP.\\
            \hline
	\end{tabular}	
	\caption{\textbf{\textcolor{black}{Some legal recommendations and / or questions posed to e-commerce platforms and our relevant observations. Positioning in SERP interfaces and interleaving sponsored results may have subtle effects on customers.}}}
	\label{Tab: LegalQueries}
\end{table*}

\noindent \textbf{Summary of results}: 
Our audit study on Amazon and Flipkart shows that Amazon advertises a significant amount of private label (PL) products ($\approx 15\%$) on the SERPs, specially on the first page (Table~\ref{tab:pl_in_ad}). The numbers for Flipkart are comparatively lower ($4.5\%$). From Figure~\ref{fig:pl_pos_ad_pos}, we observe that Amazon does most of its PL promotion on the top, middle and bottom rows of the grid-based SERP. Flipkart does the same in the top two results and the rightmost column of the SERP. These product positions also coincide with existing user attention patterns proposed in the literature. Moreover, our survey using a mock shopping website shows that users are more prone to following the middle bias attention pattern as opposed to the slower decay attention pattern. Both the platforms seem to exploit this knowledge of the user attention being middle-biased to place their PL products so that these products garner the maximum attention. Figure~\ref{fig:rand_orig} shows that Amazon benefits from its placement of the PL products on the advertised rows remarkably as compared to a random arrangement, whereas Flipkart's PL product placement strategy performs similar to a random PL product placement strategy.

\noindent
\textbf{Bridging the gap between legal principles and their opperationalisation:}
While new digital acts e.g., Digital Services Act~\cite{EC2022DSARegulation} and Digital Markets Act~\cite{EC2022DMARegulation} are already in force, the gap between the regulations and how organizations and auditors will operationalise them still exists. The methods adopted in this work can be seen as attempts to bridge these gaps and answer some of the questions systematically. Table~\ref{Tab: LegalQueries} shows some key recommendations of DMA and a question from US antitrust hearing and some insights drawn from the current study in their contexts.  

However, readers are cautioned that the analyses performed in this work are in the context of Indian marketplaces. At the same time, since the design choices adopted by global platforms e.g., Amazon are consistent across the globe, some of these methodologies can be utilised in other marketplaces as well. We by no means try to refute the answer given by Amazon to US antitrust subcommittee since our work does not have access to the statistics based on all possible web-pages and queries on Amazon. Hence the readers are cautioned to perceive these insights only in the context of the queries and dataset analysed in this study.

\noindent \textbf{Takeaways for different stakeholders}: As mentioned above, the methodologies adopted in this work can broadly be of interest to both platforms and regulators for establishing compliance / non-compliance for some of the legal recommendations (as shown in Table~\ref{Tab: LegalQueries}). 
Specific details of the methods may change, e.g., instead of a survey on AMT, e-commerce platforms can conduct A/B testings across their customer-base to understand and quantify the efficacy and compliance of their different product placement strategies regularly. 
Historically, `\textit{customer benefits}' has been the primary defense strategy of e-commerce platforms in anti-trust investigations~\cite{amz_anti_trust}. However, when the implications of these different design choices are evaluated in the perspective of different platforms -- and especially while keeping customers in the loop -- it often unfolds different dimensions of the narrative. 
For example, we observe that the ad and private label positioning strategy of Amazon is not only significantly more effective in drawing clicks than that of a random SERP, but also seems to be more effective than the strategy adopted by Flipkart. 
Further, based on the survey observations, one can conclude that customers may not have any significant preference toward private label products either; they are rather are more influenced by the positioning of these products on the SERP.

Thus not only algorithms, but also design choices (e.g., ad placement) adopted across the different platforms need to be considered for understanding which one is providing a comparatively fair competition on its marketplace or better utility for customers. For customers, the primary takeaway is to be more judicious and cautious while interacting with different SERP interfaces. 
Similarly, given e-commerce platforms are adopting such business models -- these findings may also motivate sellers to improve their business strategies in terms of investing in promoting their products, selecting platforms for conducting their business, etc.  

\noindent \textbf{Broader perspective}:
Audit studies, such as ours, can lead to a better understanding of the algorithms deployed on these platforms and their effects on all stakeholders. First, these platforms are highly complex and a single audit cannot evaluate the entire platform. Hence, individual audits like ours, are needed to comprehensively evaluate the different algorithms for a platform. Next, our study has shown that platforms exercise huge power in terms of nudging customers toward certain products of their choice by placing these in a certain way, irrespective of how the organic results are ranked or displayed. 

\textcolor{black}{Note that, in this study, our goal is \textit{not} to compare between Amazon and Flipkart. We allude to the fact that even platforms designed for the same task (e-commerce) have significant differences in how they approach placement of PL products and advertising positions. These differences are not just superficial, but may lead to deeper concerns surrounding biases.}

Our audit will make the platform developers aware of their nudging actions as well as inform policy makers about private label (PL) product exposure, not just in terms of quantity of products but also their placement. Our methodology and measurement metrics are simple and succinct, hence easily reproducible and can be used by other third-party auditors to audit similar platforms.

\bibliographystyle{IEEEtran}
\bibliography{main}

\end{document}